\documentclass[nofootinbib,twocolumn,aps,letterpaper,superscriptaddress,showpacs]{revtex4}
\usepackage{amsmath}
\usepackage{amsfonts}
\usepackage[dvips]{graphicx}
\begin{document}

\newcommand\be{\begin{equation}}
\newcommand\ee{\end{equation}}
\newcommand\bea{\begin{eqnarray}}
\newcommand\eea{\end{eqnarray}}
\newcommand\bseq{\begin{subequations}} 
\newcommand\eseq{\end{subequations}}
\newcommand\bcas{\begin{cases}}
\newcommand\ecas{\end{cases}}
\newcommand{\p}{\partial}
\newcommand{\f}{\frac}

\title{Modification of Heisenberg uncertainty relations in\\ non-commutative Snyder space-time geometry}

\author{Marco Valerio Battisti}
\email{battisti@icra.it}
\affiliation{ICRA - International Center for Relativistic Astrophysics}
\affiliation{Dipartimento di Fisica (G9), Universit\`a di Roma ``Sapienza'' P.le A. Moro 5, 00185 Rome, Italy}
\author{Stjepan Meljanac}
\email{meljanac@irb.hr}
\affiliation{Rudjer Bovskovic Institute, Bijenivcka c.54, HR-10002 Zagreb, Croatia}


\begin{abstract}
We show that the Euclidean Snyder non-commutative space implies infinitely many different physical predictions. The distinct frameworks are specified by generalized uncertainty relations underlying deformed Heisenberg algebras. Considering the one-dimensional case in the minisuperspace arena, the bouncing Universe dynamics of loop quantum cosmology can be recovered.
\end{abstract}


\maketitle 

\section{Introduction}

Non-commutative geometries are widely considered as plausible candidates for describing physics at the Planck scale \cite{noncom} and have natural connections with string theory \cite{SW}. Moreover some of these models can be related to the intuitions of doubly special relativity (DSR) \cite{AMS} where another invariant scale (apart from the speed of light) is introduced ab initio in the theory. Interest in DSR is also increased because such a framework can be regarded as a semi-classical limit of quantum gravity (see \cite{RovSmo} and references therein).

In this paper the Snyder proposal \cite{Sny} of a non-commutative space-time is analyzed from a physical point of view. This model can be understood by means of the projective geometry approach to the de Sitter space of momenta with two universal constants and is relevant since it can be related to some of DSR models \cite{Kov}. Furthermore, it has some motivations from loop quantum gravity \cite{LO} and two-time physics \cite{tt}.

The starting point of our analysis is the requirement that the only deformed commutator in the Euclidean Snyder framework is one between the coordinates. This way, the translation group is not deformed and the rotational symmetry is preserved. We then show that, infinitely many commutators between the non-commutative coordinates and momenta are possible, such that in all the cases the algebra closes. This way, infinitely many different physical predictions of the Snyder space are allowed. These are summarized in the deformed symplectic geometry and in the generalized uncertainty principle at classical and quantum level, respectively. The physical interesting framework of a deformed quantum cosmology is also analyzed. Here we deal with a one-dimensional system and our picture is almost uniquely fixed. We show that this framework naturally leads to the non-singular (bouncing) Friedmann dynamics obtained in recent issues of loop quantum cosmology (LQC) \cite{bloop}.

The paper is organized as follows. In Section II the algebraic structure of the Euclidean Snyder space is analyzed. Section III is devoted to discuss the physical implications of this framework. Concluding remarks follow. Over the paper we adopt units such that $\hbar=c=1$.

\section{Realizations of Snyder space}

The algebraic structure of the non-commutative Snyder space is analyzed in this Section. All possible realizations of this space, the general form of the uncertainty principle and the required hermiticity conditions are showed. The known algebras are then recovered as particular cases of our construction.

{\it Realizations.} Let us start by considering a $n$-di\-men\-sio\-nal non-commutative (deformed) Euclidean space such that the commutator between the coordinates has the non-trivial structure ($\{i,j,...\}\in\{1,...,n\}$)
\be\label{snyalg}
[\tilde x_i,\tilde x_j]=\kappa M_{ij}\,,
\ee
where with $\tilde x_i$ we refer to the non-commutative coordinates and $\kappa\in\mathbb R$ is the deformation parameter with dimension of a squared length. We then demand that the rotation generators $M_{ij}=-M_{ji}=i(x_ip_j-x_jp_i)$ satisfy the ordinary $SO(n)$ algebra
\be
[M_{ij},M_{kl}]=\delta_{jk}M_{il}-\delta_{ik}M_{jl}-\delta_{jl}M_{ik}+\delta_{il}M_{jk}
\ee
and that the translation group is not deformed, i.e. $[p_i,p_j]=0$. In order to preserve the rotational symmetry the commutators between $M_{ij}$ and the coordinates $\tilde x_i$, as well as between $M_{ij}$ and $p_k$, have to be undeformed. Therefore, we assume that the relations
\bea\label{commx}
[M_{ij},\tilde x_k]&=&\tilde x_i\delta_{jk}-\tilde x_j\delta_{ik}, \\\nonumber
[M_{ij},p_k]&=&p_i\delta_{jk}-p_j\delta_{ik}
\eea 
hold. This way we deal with the (Euclidean) Snyder space \cite{Sny}. The above relations however do not uniquely fix the commutators between $\tilde x_i$ and $p_j$. In particular, there are infinitely many of such commutators which are all compatible (in the sense that the algebra closes in virtue of the Jacobi identities) with the above natural requirements. 

This feature can be understood by analyzing the realizations \cite{Mel,Luk,Gosh} of such a non-commutative space. The concept of realization was developed in a series of papers \cite{Mel} (for a similar approach in the $\kappa$-deformed space-time see \cite{Luk} and a related analysis in the context of DSR can be found in \cite{Gosh}). A realization of the Snyder algebra (\ref{snyalg}) is defined as a rescaling of the non-commutative coordinates $\tilde x_i$ in terms of the ordinary phase space variables ($x_i,p_j$). The most general $SO(n)$ covariant realization for $\tilde x_i$ is given by
\be\label{real}
\tilde x_i=x_i\varphi_1(\mu,\nu)+\kappa(x_jp_j)p_i\varphi_2(\mu,\nu),
\ee
where the convention $a_ib_i=\sum_i a_ib_i$ is adopted and $\varphi_1$ and $\varphi_2$ are two arbitrary finite functions depending on the dimensionless quantities $\mu=\kappa p^2$ and $\nu=\kappa m^2$. In particular, the second quantity accounts for a mass-like term $m^2$ which can be positive, negative or zero. In order to recover the ordinary Heisenberg algebra, suitable boundary conditions on these functions have to be imposed. We have to demand that, in the $\kappa\rightarrow0$ ($\mu,\nu\rightarrow0$) limit, $\varphi_1(0,0)=1$.  

The realization above is, of course, not completely arbitrary since it depends on the adopted algebraic structure. In particular, the two functions $\varphi_1$ and $\varphi_2$ are constrained by the relations (\ref{snyalg}) and (\ref{commx}). Inserting the formula (\ref{real}) into the non-commutative coordinate commutator (\ref{snyalg}), the first restriction we obtain reads
\be\label{con1}
2\left(\varphi_1'\varphi_1+\mu\varphi_1'\varphi_2\right)-\varphi_1\varphi_2+1=0,
\ee
where $\varphi_1'=\p\varphi_1/\p\mu$. The other condition on $\varphi_1$ and $\varphi_2$ arises after considering the realization (\ref{real}) into the commutator $[M_{ij},\tilde x_k]$ in (\ref{commx}). Such second constraint can be written as
\be\label{con2}
\left(x_l[M_{ij},p_l]p_k+[M_{ij},x_l]p_lp_k\right)\varphi_2=0
\ee
and is immediate to verify that the argument in the brackets identically vanishes. Therefore, only one condition on $\varphi_1$ and $\varphi_2$ appears. As a matter of fact, given any function $\varphi_1(\mu,\nu)$ satisfying the boundary condition $\varphi_1(0,0)=1$, the function $\varphi_2(\mu,\nu)$ is uniquely determined by the equation (\ref{con1}) and reads $\varphi_2=(1+2\varphi_1'\varphi_1)/(\varphi_1-2\mu\varphi_1')$. In other words, there are infinitely many ways to express, via $\varphi_1$, the non-commutative coordinates (\ref{snyalg}) in terms of the ordinary ones without deforming either the rotation and the translation groups.

It is worth noting that: (i) The realizations (\ref{real}) have sense if there exist the inverse transformation $x_i=\tilde x_j(\varphi^{-1})_{ji}$ and the necessary and sufficient condition is $\det|\delta_{ij}\varphi_1+\kappa p_ip_j\varphi_2|>0$. If we deal with a $n\geq2$ dimensional space, such a condition reads $\varphi^{n-1}_1(\varphi_1+\mu\varphi_2)>0$, i.e. $\varphi_1>0$ and $\varphi_1+\mu\varphi_2>0$. (ii) Our analysis can be straightforward generalized to a Snyder Minkowskian space-time. In this case, all the relations above hold as soon as the following replacements are taken into account ($\{\alpha,\beta\}\in\{0,...,n\}$): the $SO(n)$ generators are substituted by the Lorentz generators $L_{\alpha\beta}$ and $(\tilde x_\alpha,p_\alpha)$ now transform as four-vectors under Lorentz algebra which indices are raised and lowered by the Minkowski metric $\eta_{\alpha\beta}$, i.e  $p^2=\eta^{\alpha\beta}p_\alpha p_\beta$ is Lorentz invariant.   

{\it Uncertainty relations.} In order to complete the analysis of the deformed algebra we need to analyze the $(\tilde x-p)$ commutation relation. This way, the general form of the uncertainty principle, and thus the physical consequences of the model, can be discussed. The commutator between $\tilde x_i$ and $p_j$ arises from the realization (\ref{real}) and reads
\be\label{xpcom}
[\tilde x_i,p_j]=i\left(\delta_{ij}\varphi_1+\kappa p_ip_j\varphi_2\right).
\ee
Of course, the ordinary one is recovered in the $\kappa\rightarrow0$ limit. From the commutator above we can immediately obtain the generalized uncertainty principle underlying the Snyder non-commutative space, i.e.
\be\label{uncrel}
\Delta\tilde x_i\Delta p_j\geq\f12|\delta_{ij}\langle\varphi_1\rangle+\kappa\langle p_ip_j\varphi_2\rangle|.
\ee
Three remarks are in order. (i) The algebra we obtain can be regarded as a deformed Heisenberg algebra. More precisely, the deformation of the only commutator between the spatial coordinates as in (\ref{snyalg}) leads to infinitely many realizations of the algebra, and thus generalized uncertainty relations (\ref{uncrel}), all consistent with the assumptions underlying the model. (ii) Unless $\varphi_2=0$ no compatible observables exist. These are coupled with each other and an exactly simultaneous measurable couple $(\tilde x_i,p_j)$ is not longer allowed. A measure of the $i$-component of the (non-commutative) position will always affect a measure of the $j(\neq i)$-component of the momentum by an uncertainty $\Delta p_j\gtrsim|\kappa\langle p_ip_j\varphi_2\rangle|/\Delta\tilde x_i$. (iii) For any fixed $\varphi_1$ the non-commutative framework is unique, but we can realize the commutator (\ref{xpcom}) in terms of any commutative coordinates $x_i'$ and corresponding canonical momenta $p_i'$ satisfying $[x_i',p_j']=i\delta_{ij}$. Of course all these descriptions lead to the same physical consequences.  

{\it Hermiticity conditions.} The non-commutative coordinates $\tilde x_i$ have to be hermitian operators in any given realization. All the commutators given above are invariant under the formal anti-linear involution ``$\dag$''
\be
\tilde x_i^\dag=\tilde x_i, \quad p_i^\dag=p_i, \quad M_{ij}^\dag=-M_{ij}\,,
\ee
where the order of elements is inverted under the involution. However, the realization (\ref{real}) in general is not hermitian. The hermiticity condition can be immediately implemented as soon as the expression
\be
\tilde x_i=\f12\left(x_i\varphi_1+\kappa(x_jp_j)p_i\varphi_2+\varphi_1^\dag x_i^\dag+\kappa\varphi_2^\dag p_i^\dag(x_jp_j)^\dag\right)
\ee 
is taken into account. However, the physical result do not depend on the choice of the representation as long as exist a smooth limit $\tilde x_i\rightarrow x_i$ as $\kappa\rightarrow0$. Therefore, we can restrict our attention to non-hermitian realization only.

{\it Recovering the know realizations.} The non-com\-mu\-ta\-ti\-ve Snyder space has been analyzed in literature from different points of view \cite{Kov,LO,tt} (see also \cite{GB}), but only two particular realizations of its algebra are known: the Snyder \cite{Sny} and the Maggiore \cite{Mag} ones. The original realization of Snyder, in particular, expressed through the commutator between $\tilde x$ and $p$, reads
\be\label{xpsny}
[\tilde x_i,p_j]=i\left(\delta_{ij}+\kappa p_ip_j\right).
\ee
It is not difficult to see that this is a particular case of our realization (\ref{real}) as soon as $\varphi_1=1$. From this condition, the function $\varphi_2$ is fixed by (\ref{con1}) as $\varphi_2=1$ and the above commutation relation is recovered. The condition on the inverse mapping implies that $p^2>-1/\kappa$. On the other hand, the Maggiore algebra 
\be\label{magalg}
[\tilde x_i,p_j]=i\delta_{ij}\sqrt{1-\kappa(p^2+m^2)},
\ee
can be regarded as the particular case of (\ref{real}) when the condition $\varphi_2=0$ is taken into account. But this requirement alone is not enough. In fact, from the constraint (\ref{con1}), the function $\varphi_1$ is not uniquely fixed but reads $\varphi_1=\sqrt{1-\mu+f(\nu)}$, where $f(\nu)$ is a generic function of $\nu$ (the inverse mapping condition entails $p^2<(1+f)/\kappa$). Only in the specific case $f(\nu)=-\nu$ the commutator (\ref{magalg}) is recovered. Finally, we note that, the deformed algebra proposed by Kempf et al. in \cite{Kem} can be regarded as a particular case of (\ref{magalg}) as $|\mu|\ll1$ and $m=0$, i.e. $[\tilde x_i,p_j]=i\delta_{ij}(1+\beta p^2)$ where $\beta=-\kappa/2$ with $\kappa<0$. In the one-dimensional framework (see below), this algebra is the same of the Snyder one (\ref{xpsny}).  

\section{Physical implications} 

As understood, the physical consequences of a non-commutative space geometry are deeply and completely new scenarios are opened at both classical and quantum levels. Two physically relevant frameworks are analyzed in this Section: a generic mechanical system and the so-called quantum cosmological arena. 

{\it Mechanical system.} Let us start by considering a mechanical system, i.e. a model with a finite number of degrees of freedom described by a Hamiltonian $H=H(\tilde x,p)$. At classical level the deformations induced on the system appear as soon as the (classical) limit $-i[\cdot,\cdot]\rightarrow\{\cdot,\cdot\}$ is taken into account. In doing this the deformation parameter $\kappa$ is regarded as an independent constant with respect to $\hbar$. Modifications of a non-commutative framework on the classical dynamics are then summarized in the deformed Poisson brackets 
\be
\{F,G\}=\left(\f{\p F}{\p\tilde x_i}\f{\p G}{\p p_j}-\f{\p F}{\p p_i}\f{\p G}{\p\tilde x_j}\right)\{\tilde x_i,p_j\}+\f{\p F}{\p\tilde x_i}\f{\p G}{\p\tilde x_j}\{\tilde x_i,\tilde x_j\}
\ee  
between any phase space functions. This symplectic structure is not fixed but depends on the two functions $\varphi_1$ and $\varphi_2$ constrained by (\ref{con1}) and $\varphi_1(0,0)=1$. From the relation above, the equations of motion of a mechanical system are deformed as 
\bea\label{eqmod}
\dot{\tilde x}_i&=&\{\tilde x_i,H\}=\f{\p H}{\p p_j}\left(\delta_{ij}\varphi_1+\kappa p_ip_j\varphi_2\right)+\f\kappa i\f{\p H}{\p\tilde x_j}M_{ij}, \nonumber\\
\dot p_i&=&\{p_i,H\}=-\f{\p H}{\p\tilde x_j}\left(\delta_{ij}\varphi_1+\kappa p_ip_j\varphi_2\right). 
\eea
When the deformation parameter vanishes ($\kappa\rightarrow0$) the ordinary Hamilton equations are recovered. At quantum level our picture implies either modifications of the Ehrenfest theorem through (\ref{eqmod}), either deformations of the canonical quantization prescription via the commutator (\ref{xpcom}). As we said, this commutator is not fixed at all by the assumptions described above and for any choice of the realization (\ref{real}) of the non-commutative coordinates, the corresponding Hilbert spaces are thus unitarily inequivalent. Each quantum framework (Hilbert space) corresponds to a specific choice of the realization (\ref{real}). We also stress that given an eigenvalue problem $\hat H(\tilde x,p)\psi(x)=E\psi(x)$, the wave function $\psi$ and the spectrum $E$ depend on $\varphi_1$.

This is not surprising since the deformation of the canonical commutation relations can be viewed, from the realization (\ref{real}), as an algebra homomorphism which is a non-canonical transformation. In particular, it can not be implemented at quantum level as an unitary transformation. From this point of view, the set of predictions of any deformed Heisenberg algebra can not be matched by the set of predictions arising from another one, i.e. neither by the set of prediction of the ordinary framework (the $\kappa\rightarrow0$ limit). New features are then introduced, for any fixed $\varphi_1$, at both classical and quantum level. This way, a Snyder structure (\ref{snyalg}) in which the translation and rotation groups are undeformed, leads to infinitely many different physical predictions through (\ref{real}). 

A notable problem to be considered is the harmonic oscillator with non-commutative quadratic potential, i.e. $H=p^2/2m+m\omega^2\tilde x^2/2$. In the one-dimensional case the symmetry group is trivial ($SO(1)=\text{Id}$) and the most general realization is given by $\tilde x=x\sqrt{1-\mu+f(\nu)}$. Considering the representation of this algebra (we take $f=0$) in the momentum space, the deformed stationary Schr\"odinger equation for this model is given by the so-called Mathieu equation and the energy spectrum appears to be modified as $E_n=\omega(2n+1)/2-\omega\kappa(2n^2+2n+1)/8d^2+\mathcal O(\kappa^2/d^4)$ where $d=1/\sqrt{m\omega}$ is the characteristic length scale (for more details see \cite{Bat}). We note that, if $\kappa>0$ this is the spectrum obtained in polymer (loop) quantum mechanics \cite{pol}, while if $\kappa<0$ this result resembles the one recovered in \cite{DJM03}. 

{\it Quantum cosmology.} An interesting quantum mechanical arena to test such a framework is the so-called minisuperspace reduction of the dynamics, i.e. quantum cosmology. As well-know \cite{Wald}, by imposing symmetries on the spatial Cauchy surfaces which fill the space-time manifold, a considerable simplification on the gravitational theory occurs. In particular, by requiring the spatial homogeneity the phase space of general relativity reduces to six dimensions. The system is described by three degrees of freedom, i.e. the three scalar factors of the Bianchi models. Furthermore, by imposing also the spatial isotropy, we deal with one-dimensional mechanical systems. These are the Friedmann-Robertson-Walker (FRW) models which describe the observed Universe and on which the standard model of cosmology is based. 

In order to discuss the implications of the Snyder algebra on the FRW Universes we consider the one-dimensional case of the scheme analyzed above. If we assume the minisuperspace as Snyder-deformed, then the isotropic scale factor $a$ (namely the radius of the Universe) and its conjugate momentum $p$ satisfy the commutation relation $[a,p]=i\sqrt{1-\mu+f(\nu)}$. It is worth stressing that, when $\kappa>0$ (taking $f=0$) a natural cut-off on the momentum arises, i.e. $|p|<\sqrt{1/\kappa}$, while as $\kappa<0$ the uncertainty relation (\ref{uncrel}) predicts a minimal observable length $\Delta{\tilde x}_\text{min}=\sqrt{-\kappa}$. Moreover, at the first order in $\kappa$, the string theory result \cite{String} $\Delta{\tilde x}\gtrsim(1/\Delta p+l_s^2\Delta p)$, in which the string length $l_s$ can be identify with $\sqrt{-\kappa/2}$, is recovered. 

Following \cite{Bat} is possible to show that the effective Friedmann equation of Snyder-deformed flat FRW cosmological model becomes 
\be\label{modfri}
\left(\f{\dot a}a\right)^2=\f{8\pi G}3\rho\left(1-\f\rho{\rho_c}+f(\nu)\right),
\ee
where $G$ is the gravitational constant, $\rho=\rho(a)$ denotes a generic matter energy density and $\rho_c=(2\pi G/3\kappa)\rho_P$ is the critical energy density ($\rho_P$ being the Planck one). When the limit $\kappa\rightarrow0$ is taken into account, the critical energy density diverges (the function $f(\nu)$ disappears) leading to the ordinary dynamics. It is worth noting that, if $f(\nu)=0$ and $\kappa>0$, the equation (\ref{modfri}) resembles exactly the effective bouncing Friedmann equation of LQC \cite{bloop}. Such a dynamics is singularity-free since, when $\rho$ reaches the critical energy density, $\dot a$ vanishes and the Universe experiences a (big)-bounce instead of the classical big-bang. On the other hand, if $f(\nu)=0$ and $\kappa<0$, the effective braneworlds dynamics is recovered \cite{Roy}. 

Summarizing, the non-commutative Snyder minisuperspace framework can clarify similarities and differences between different quantum gravity theories. Other comparisons between deformed and loop-polymer quantum cosmology, in view of discussing the fate of the cosmological singularity at quantum level, were performed considering the flat FRW model filled with a massless scalar field \cite{BM07a} and the Taub cosmological model \cite{BM07b}. Such investigations deserve interest either in clarifying the role of loop quantization techniques in cosmology, either in establishing a phenomenological contact with some frameworks relevant in a flat space-time limit of quantum gravity. 

\section{Concluding remarks}

In this paper we have shown how there are infinitely many realizations of the Snyder algebra, equations (\ref{snyalg}-\ref{commx}), implying different commutation relations between the non-commutative coordinates $\tilde x$ and momenta $p$, i.e. we deal with deformed Heisenberg algebras. These depend on an arbitrary function $\varphi_1(\mu,\nu)$ such that $\varphi_1(0,0)=1$ ensuring the correctness of the picture. Therefore, different non-commutative spaces, described by distinct commutations relations (\ref{xpcom}), imply different (unitarily inequivalent) physical consequences. On the other hand, in the one-dimensional case the commutator between $\tilde x$ and $p$ is fixed (up to a function of the mass-like term) and, when implemented in the minisuperspace dynamics, the loop as well as the braneworlds cosmological evolutions are recovered. 

{\it Acknowledgments.} We thank Daniel Meljanac for comments. M.V.B. thanks S.M. for the warm hospitality in Zagreb during which this paper was carried out. This work was supported in part by Ministry of Science and Technology of the Republic of Croatia under contract No. 098-0000000-2865.


\begin{thebibliography}{99}

\bibitem{noncom}M. R. Douglas and N. A. Nekrasov, Rev.Mod.Phys. 73 (2001) 977; S. Doplicher, K. Fredenhagen and J. E. Roberts, Phys.Lett.B 331 (1994) 39.

\bibitem{SW}N. Seiberg and E. Witten, JHEP 9909 (1999) 032. 

\bibitem{AMS}G. Amelino-Camelia, Int.J.Mod.Phys.D 11 (2002) 35; Phys.Lett.B 510 (2001) 255; J. Magueijo and L. Smolin, Phys.Rev.Lett. 88 (2002) 190403.

\bibitem{RovSmo}C. Rovelli, arXiv:0808.3505; L. Smolin, arXiv:0808.3765.

\bibitem{Sny}H. S. Snyder, Phys.Rev. 71 (1947) 38.

\bibitem{Kov}J. Kowalski-Glikman, Phys.Lett.B 547 (2002) 291; J. Kowalski-Glikman and S. Nowak, Class.Quant.Grav. 20 (2003) 4799; H. Guo, C. Huang and H. Wu, Phys.Lett.B 663 (2008) 270.

\bibitem{LO}E. R. Livine and D. Oriti, JHEP 0406 (2004) 050.

\bibitem{tt}J. M. Romero and A. Zamora, Phys.Rev.D 70 (2004) 105006.  

\bibitem{bloop}A. Ashtekar, T. Pawlowski and P. Singh, Phys.Rev.D 73 (2006) 124038; P. Singh, Phys.Rev.D 73 (2006) 063508.

\bibitem{Mel}L. Jonke and S. Meljanac, Phys.Lett.B 526 (2002) 149; S. Meljanac and M. Stojic, Eur.Phys.J.C 47 (2006) 531; S. Kresic-Juric, S. Meljanac and M. Stojic, Eur.Phys.J.C 51 (2007) 229; T. R. Govindarajan, K. S. Gupta, E. Harikumar, S. Meljanac and D. Meljanac, Phys.Rev.D 77 (2008) 105010. 

\bibitem{Luk}J. Lukierski, H. Ruegg and W. J. Zakrzewski, Annals.Phys. 243 (1995) 90.

\bibitem{Gosh}S. Ghosh and P. Pal, Phys.Rev.D 75 (2007) 105021.

\bibitem{GB}R. Banerjee, S. Kulkarni and S. Samanta, JHEP 0605 (2006) 077; L. A. Glinka, arXiv:0812.0551.

\bibitem{Mag}M. Maggiore, Phys.Lett.B 304 (1993) 65; Phys.Rev.D 49 (1994) 5182.

\bibitem{Kem}A. Kempf, G. Mangano and R. B. Mann, Phys.Rev.D 52 (1995) 1108; A. Kempf, J.Math.Phys. 38 (1997) 1347.

\bibitem{Bat}M. V. Battisti, arXiv:0805.1178; J.Phys.Conf.Ser. (2008) at press, arXiv:0810.5039.

\bibitem{pol}A. Ashtekar, S. Fairhurst and J. L. Willis, Class.Quant.\\Grav. 20 (2003) 1031.

\bibitem{DJM03}I. Dadic, L. Jonke and S. Meljanac, Phys.Rev.D 67 (2003) 087701.

\bibitem{Wald}R. M. Wald, {\it General Relativity} (CUP, Chicago, 1984).

\bibitem{String}D. J. Gross and P. F. Mendle, Nucl.Phys.B 303 (1988) 407; K. Konishi, G. Paffuti and P. Provero, Phys.Lett.B 234 (1990) 276.

\bibitem{Roy}R. Maartens, Living Rev.Rel. 7 (2004) 7.

\bibitem{BM07a}M. V. Battisti and G. Montani, Phys.Lett.B 656 (2007) 96

\bibitem{BM07b}M. V. Battisti and G. Montani, Phys.Rev.D 77 (2008) 023518; M. V. Battisti, O. M. Lecian and G. Montani  Phys.Rev.D 78 (2008) 103514. 

\end{thebibliography}
\end{document}